\documentclass[prb,preprint]{revtex4-1}   

\usepackage{amsmath}    
\usepackage{amsfonts}  
\usepackage{graphicx}   
\usepackage{bm}

\newcommand{\ket}[1]{{|{#1}\!\!>}}
\newcommand{\spi}[2]{{\left[ \begin{array}{c} {#1} \\ {#2} \end{array} \right]}}
\newcommand{\fspi}[4]{{\left[ \begin{array}{c} {#1} \\ {#2} \\ {#3} \\ {#4} \end{array} \right]}}
\begin{document}

\title{A classical analog for the electron spin state}
\author{K.B. Wharton, R.A. Linck and C.H. Salazar-Lazaro}

 \affiliation{San Jos\'e State University, Department of Physics and Astronomy, San Jos\'e, CA 95192-0106}
 \email{wharton@science.sjsu.edu}   

\date{\today}

\begin{abstract}

Despite conventional wisdom that spin-1/2 systems have no classical analog, we introduce a set of classical coupled oscillators with solutions that exactly map onto the dynamics of an unmeasured electron spin state in an arbitrary, time-varying, magnetic field.  While not addressing the quantum measurement problem (discrete outcomes and their associated probabilities), this new classical analog yields a classical, physical interpretation of Zeeman splitting, geometric phase, the electron's doubled gyromagnetic ratio, and other quantum phenomena.  This Lagrangian-based model can be used to clarify the division between classical and quantum systems, and might also serve as a guidepost for certain approaches to quantum foundations.

\end{abstract}

\maketitle

\section{Introduction}

%
%

Despite the conventional view of quantum spin as being an inherently non-classical phenomenon\cite{LL}, there is a rich history of exploring classical analogs for spin-1/2 systems in particular.  For example, there exists a well-developed classical analog to a two-level quantum system, based upon the classical polarization (CP) of a plane electromagnetic wave\cite{Mcmaster,HnS,Klyshko,Malykin,Zap}.  Although this CP-analog has been used to motivate introductory quantum mechanics texts\cite{Baym, Sakurai}, the power and depth of the analogy is not widely appreciated.  For example, the CP-analog contains a straightforward classical picture for a $\pi$ geometric phase shift resulting from a full $2\pi$ rotation of the spin angular momentum, but this fact is rarely given more than a casual mention (with at least one notable exception\cite{Klyshko}).  Still, the CP-analog contains certain drawbacks, especially when the analogy is applied to an electron spin state in an arbitrary, time-varying magnetic field.  These drawbacks, along with complications involving quantum measurement outcomes, have prevented a general agreement on exactly which aspects of quantum spin are inherently non-classical.

In this paper, we extend the CP-analog to a system of four coupled oscillators, and prove that this classical system exactly reproduces the quantum dynamics of an unmeasured electron spin state in an arbitrary magnetic field.  This result demonstrates, by explicit construction, that if there are any aspects of an electron spin state that cannot be described in a classical context, those aspects must lie entirely in the domain of quantum measurement theory, not the dynamics.  In order to accomplish this feat, it turns out there must necessarily be a many-to-one map from the classical system to the quantum state.  In other words, the classical system contains a natural set of ``hidden variables'', accessible to the classical analog, but hidden to a complete specification of the quantum state.

Some might argue that no classical analog is needed to discuss quantum spin dynamics because an unmeasured quantum state governed by the Schr\"odinger-Pauli Equation (SPE) could be interpreted as a set of classical quantities coupled by first-order differential equations.  One can even analyze the classical Dirac field and deduce quantities which map nicely onto quantum spin concepts \cite{Ohanian}.  But simply reinterpreting quantum wave equations as classical fields is not a very enlightening ``analog'', especially if the spin state is considered separately from the spatial state.  For example, the use of complex numbers in these equations is significantly different from how they are used to encode phases in classical physics, and therefore have no obvious classical interpretation.  And if a system of first-order differential equations cannot be directly transformed into a set of second-order differential equations, it is unclear how certain classical physics concepts (\textit{e.g.} generalized forces) can be applied.  As we will demonstrate below, the full SPE \textit{can} be expanded to a system of second-order equations, but only by adding additional ``hidden variables'' along with new overall constraints.  The classical analog presented here arrives at this result from a different direction, starting with a simple Lagrangian.

Apart from clarifying how quantum spin might best be presented to students, the question of which aspects of quantum theory are truly ``non-classical'' is of deep importance for framing our understanding of quantum foundations.  For example, Spekkens has recently demonstrated a simple classical toy theory that very roughly maps onto two-level quantum systems, showing several examples of purportedly-quantum phenomena that have a strong classical analog\cite{Spekkens}.  Still, neither Spekkens nor other prominent classical-hidden-variable approaches to two-level quantum systems\cite{Bell, KS} have concerned themselves with classical analogies to the curious \textit{dynamics} of such systems.  Our result demonstrates that starting with the dynamics can naturally motivate particular foundational approaches, such as a natural hidden variable space on which classical analogies to quantum measurement theory might be pursued.  And because this classical analog derives from a simple Lagrangian, it is potentially a useful test bed for approaches where the action governs the transition probabilities, as in quantum field theory.

The plan for the paper is as follows:  After summarizing the CP-analog in Section II, a related two-oscillator analog (similar to a Foucault Pendulum) is presented in Section III.  This two-oscillator analog is shown to be identical to a quantum spin state in a one-dimensional magnetic field; a three-dimensional field requires an extension to four oscillators, as shown in Section IV.  The final section discusses and summarizes these results -- the most notable of which is that a classical system can encompass all the dynamics of a quantum spin-1/2 state.

\section{The Classical Polarization Analog}

For a classical plane electromagnetic (EM) wave moving in the z-direction with frequency $\omega_0$, the transverse electric fields $E_x(t)$ and $E_y(t)$ in the $z=0$ plane can always be expressed in two-vector notation as the real part of 
\begin{equation}
\label{eq:cp}
\spi{E_x}{E_y}= \spi{a}{b} e^{i \omega_0 t}.
\end{equation}
Here $a$ and $b$ are complex coefficients, encoding the amplitude and phase of two orthogonal polarizations.  

A strong analogy can be drawn between the two-vector $(a,b)$ on the right side of (\ref{eq:cp}) -- the well-known ``Jones vector" -- and the spinor $\ket{\chi}$ that defines a spin-1/2 state in quantum mechanics.  The quantum normalization condition $<\!\!\chi |\chi \!\!>=1$ maps to a normalization of the energy density of the EM wave, and the global phase transformation $\ket{\chi} \to \ket{\chi} \, exp(i \theta)$ is analogous to changing the EM wave's phase. 

This equivalence between a spinor and a Jones vector can be made more explicit by projecting them both onto the surface of a unit sphere in an abstract space (the ``Bloch sphere'' and the ``Poincar\'e sphere'' respectively).  Each spinor/Jones vector maps to a unit vector in the angular direction $(\theta, \phi)$, according to the usual convention $a=cos(\theta/2)$, $b=sin(\theta/2)exp(i\phi)$.  This is more familiarly described in terms of the sphere's six intersections with a centered cartesian axis
\begin{align}
\label{eq:defs}
\ket{z_+} &= \spi{1}{0} & \ket{z_-} &= \spi{0}{1} \notag \\
 \ket {x_+} &= \frac{1}{\sqrt{2}} \spi{1}{1} & \ket{x_-} &= \frac{1}{\sqrt{2}} \spi{1}{-1} \\
 \ket{y_+} &= \frac{1}{\sqrt{2}} \spi{1}{i} & \ket{y_-} &= \frac{1}{\sqrt{2}} \spi{1}{-i}. \notag
\end{align}

The CP-analog therefore maps linear x-polarized light to a spin-up electron $\ket{z_+}$ and linear y-polarized light to a spin-down electron $\ket{z_-}$.  Electrons with measured spins $\ket{x_\pm}$ correspond to xy-diagonal linear polarizations, while $\ket{y_\pm}$ correspond to the two circular-polarization modes.  In this framework, note that quantum superpositions are no different than ordinary EM superpositions.

The analogy extends further, but this is already sufficient background to classically motivate some of the strange features of spin-1/2 systems.  Consider the rotation of a Jones vector around the equator of a Poincar\'e sphere, corresponding to a continuous rotation of the direction of linear polarization -- from horizontal, through vertical, and back to the original horizontal state.  Any transformation that leads to this rotation (say, a physical rotation of the wave medium) will then be analogous to a magnetic-field induced precession of a spin state around the corresponding equator of the Bloch sphere.

The key point is that the above-described $2\pi$ rotation around the Poincar\'e sphere merely corresponds to a $\pi$ rotation of the EM polarization in physical space.  And this is equivalent to a $\pi$ phase shift in the resulting wave; it would now interfere destructively with an identical unrotated wave.  Of course, this is also the observed effect for a $2\pi$ rotation of a quantum spin state around the Bloch sphere, although in the latter case the net geometric phase shift is generally thought to be inexplicable from a classical perspective.  

What the CP-analog accomplishes is to demonstrate that such behavior does indeed have a straightforward classical interpretation, because the geometrical phase of the spin state is directly analogous to the overall phase of the physical EM wave \cite{Klyshko}.  The key is that the Poincar\'e sphere does not map to physical space, so a $2\pi$ rotation need not return the EM wave to its original state.  The CP-analog therefore advocates the viewpoint that the Bloch sphere should not map to physical space, even for an electron spin state.  This viewpoint will be implemented below in a fully consistent fashion.

To our knowledge, it has not been explicitly noted that this classical analogy naturally motivates an apparently-doubled gyromagnetic ratio for the electron.  In the above-described Poincar\'e sphere rotation, as the EM wave is being rotated around its propagation axis, suppose an observer had reference to another system (say, a gyroscope) that truly recorded rotation in physical space.  As compared to the gyroscope, the Jones vector would seem to complete a full rotation in half the time.  If one interpreted the Poincar\'e sphere as corresponding to real space, the natural conclusion would be that the Jones vector was coupled to the physical rotation at double its ``classical'' value.  Misinterpreting the Bloch sphere as corresponding to physical space would therefore lead to exactly the same conclusion for the electron's gyromagnetic ratio. 

The classical polarization analog can be pursued much further than is summarized here, mapping the quantum dynamics induced by a magnetic field to the effects of different birefringent materials \cite{Klyshko,Malykin,Zap,Baym,Kubo}.  The two EM modes in such materials then map to the two energy eigenstates, and generic rotations around the Poincar\'e sphere can be given a physical implementation.  Still, this analog becomes quite abstract; there is no easy-to-describe vector quantity of a birefringent material that corresponds to the magnetic field, and the situation is even more convoluted for time-dependent field analogs.  

Another disanalogy is the relation between the magnitude of the Zeeman energy splitting and the difference in wavenumber of the two EM modes.  A more natural analogy would relate energy to a classical frequency, but the two EM modes always have identical frequencies.  And of course, an electromagnetic plane wave cannot be pictured as a confined system with internal spin-like properties.  In the next sections, we develop a novel classical analog that alleviates all of these problems.

\section{A Foucault-Pendulum-Like Analog}

The central success of the CP-analog stems from its use of two physical oscillators, which need not be electromagnetic.  For any two uncoupled classical oscillators with the same natural frequency $\omega_0$, their solution can also be encoded by two complex numbers $a$ and $b$, representing the amplitude and phase of each oscillator.  Therefore the use of Jones vectors and the Poincar\'e sphere does not only pertain to EM waves.

As an intermediate step towards our proposed classical analog for an electron spin state, consider this classical Lagrangian:
\begin{eqnarray}
L_1(x_1,x_2,\dot{x}_1,\dot{x}_2,t)=\frac{1}{2}\left[ p_1^2+p_2^2 - \omega_0^2(x_1^2+x_2^2)\right], \label{eq:L1}\\
\spi{p_1}{p_2} \equiv \spi{\dot{x_1}}{\dot{x_2}} + \left[ \begin{array}{cc} 0 & -\beta \\ \beta & 0 \end{array} \right] \spi{x_1}{x_2}. \label{eq:p1}
\end{eqnarray}
Here the $x's$ are all purely real quantities, and $\beta$ is some coupling constant that may be time-dependent.  (As $\beta\to 0$, this becomes two uncoupled classical oscillators).  Equation (\ref{eq:p1}) can be rewritten as $\bm{p}=\dot{\bm{x}}+\bm{B} \bm{x}$, where the conjugate momenta $p_1$ and $p_2$ form the column vector $\bm{p}$, etc.  Note that squaring the $\bm{B}$ matrix yields $\bm{B}^2=-\beta^2 \bm{I}$.  In this notation, $L_1=(\bm{p\cdot p}-\omega_0^2 \,\bm{x\cdot x})/2$.  

First, consider the case of a constant $\beta$.  The Euler-Lagrange equations of motion for $L_1$ are then
\begin{eqnarray}
\ddot{x_1}+(\omega_0^2-\beta^2) x_1 = 2\beta \dot{x}_2, \nonumber \\
\ddot{x_2}+(\omega_0^2-\beta^2) x_2 = -2\beta \dot{x}_1. \label{eq:2d1}
\end{eqnarray}
These equations happen to describe the projection of a Foucault pendulum into a horizontal plane (with orthogonal axes $x_1$ and $x_2$) in the small-angle limit.  Specifically, $\beta=\Omega sin(\phi)$, where $\Omega$ is the rotation frequency of the Earth and $\phi$ is the latitude of the pendulum.  (The natural frequency of such a pendulum is actually $\sqrt{\omega_0^2-\beta^2}$, because of a $\beta^2$ term in $L_1$ that does not appear in the Foucault  pendulum Lagrangian, but for a constant $\beta$ this is just a renormalization of $\omega_0$).

The precession of the Foucault pendulum therefore provides a qualitative way to understand the effect of a constant $\beta$ on the unnormalized Jones vector $\bm{x}$.  Given a non-zero $\beta$, it is well-known that linear oscillation in $x_1$ (mapping to $\ket{z_+}$ on the Poincar\'e sphere) precesses into a linear oscillation in $x_2$ (mapping to $\ket{z_-}$) and then back to $x_1$ ($\ket{z_+}$).  But this $2\pi$ rotation of $\bm{x}$ around the Poincar\'e sphere merely corresponds to a $\pi$ rotation of the pendulum's oscillation axis in physical space, leaving the overall phase of the pendulum shifted by $\pi$, exactly as was described for the CP-analog.

Quantitatively, solutions to (\ref{eq:2d1}) are of the form $exp(-i\omega_\pm t)$, where $\omega_\pm =\omega_0 \pm \beta$.  The generic solution can always be expressed as the real component of
\begin{equation}
\label{eq:2ds}
\spi{x_1}{x_2}= a\spi{1}{i} e^{-i (\omega_0- \beta ) t} + b\spi{1}{-i} e^{-i (\omega_0 + \beta )t}.
\end{equation}
Here $a$ and $b$ are arbitrary complex parameters (although again note that $x_1$ and $x_2$ are purely real).

One notable feature of this result is that the coupling constant $\beta$ has the effect of producing two solutions with well-defined frequencies equally spaced above and below the natural frequency -- just like Zeeman splitting of an electron's energy levels in a magnetic field.  Furthermore, the modes that correspond to these two pure frequencies happen to be right- and left-hand circular motion of the pendulum, directly analogous to $\ket{y_+}$ and $\ket{y_-}$.  A comparison of (\ref{eq:2ds}) with standard results from quantum mechanics reveals that $\beta$ produces exactly the same dynamics on $\bm{x}$ as does a constant magnetic field in the $y$ direction on an electron spin state (apart from an overall $exp(-i\omega_0 t)$ global phase).  

Given the strong analogy between a constant $\beta$ and a constant (one-component) magnetic field, one can ask whether this correspondence continues to hold for a time-varying $\beta$.  In this case the strict analogy with the Foucault pendulum fails (thanks to the $\beta^2$ terms in $L_1$) and comparing the exact solutions becomes quite difficult.  But starting from the Euler-Lagrange equations for a time-varying $\beta$,
\begin{eqnarray}
\ddot{x_1}+(\omega_0^2-\beta^2) x_1 &=& 2\beta \dot{x}_2+\dot{\beta} x_2, \nonumber \\
\ddot{x_2}+(\omega_0^2-\beta^2) x_2 &=& -2\beta \dot{x}_1-\dot{\beta} x_1, \label{eq:2d2}
\end{eqnarray}
one can compare them directly to the relevant Schr\"odinger-Pauli Equation (SPE).  Using a $y$-directed magnetic field $B_y=2\beta/\gamma$ (where $\gamma=-e/m$ is the gyromagnetic ratio) and an overall phase oscillation corresponding to a rest mass $mc^2=\hbar\omega_0$, this yields
\begin{equation}
\label{eq:sey}
i\hbar \frac{\partial}{\partial t} \spi{a}{b} = \hbar \left[ \begin{array}{cc} \omega_0 & i\beta \\ -i\beta & \omega_0 \end{array} \right] \spi{a}{b}.
\end{equation}

Taking an additional time-derivative of (\ref{eq:sey}), and simplifying the result using (\ref{eq:sey}) itself, it is possible to derive the following second-order differential equations:
\begin{eqnarray}
\ddot{a}+(\omega_0^2-\beta^2) a &=& 2\beta \dot{b}+\dot{\beta} b, \nonumber \\
\ddot{b}+(\omega_0^2-\beta^2) b &=& -2\beta \dot{a}-\dot{\beta} a. \label{eq:sey2}
\end{eqnarray}
While $a$ and $b$ are still complex, the real and imaginary parts have naturally split into separate coupled equations that are formally identical to (\ref{eq:2d2}).  So every solution to the SPE (\ref{eq:sey}) must therefore have a real component which solves (\ref{eq:2d2}).  

At first glance it may seem that the imaginary part of (\ref{eq:sey2}) contains another set of solutions not encoded in the real part of (\ref{eq:sey2}), but these solutions are not independent because they also solve (\ref{eq:sey}).  The solution space of (\ref{eq:sey}) is a complex vector space of dimension 2 over the complex numbers.  It can be verified that the SPE with a rest-mass oscillation cannot admit purely real solutions. Also, it is an elementary exercise to show that if a vector space over the complex numbers has a function basis given by $\{ z_1, z_2 \}$ and there is no complex linear combination of $z_1, z_2$ that yields a purely real function, then the set $\{ Re(z_1), Re(z_2), Im(z_1), Im(z_2) \}$ is a linearly independent set of real functions where linear independence is taken over the reals instead of the complex numbers. From this elementary result, it follows that if $\{z_1, z_2\}$ is a basis for the solution space of (\ref{eq:sey}) over the complex numbers, then the set of functions $X= \{Re(z_1), Re(z_2), Im(z_1), Im(z_2) \}$ spans a 4-d real subspace of the solution space of (\ref{eq:2d2}). Since (\ref{eq:2d2}) indeed has a 4-d solution space over the reals, it follows that the subspace spanned by $X$ is indeed the full solution space of (\ref{eq:2d2}). In summary, the solutions to the real, second-order differential equations (\ref{eq:2d2}) exactly correspond to the solutions to the complex, first-order differential equations (\ref{eq:sey}).

For a one-dimensional magnetic field, these results explicitly contradict the conventional wisdom concerning the inherent complexity of the spin-1/2 algebra.  By moving to real second-order differential equations -- a natural fit for classical systems -- it is possible to retain exactly the same dynamics, even for a time-varying magnetic field.  The resulting equations not only account for a Zeeman-like frequency splitting, but demonstrate that the quantum geometric phase can be accounted for as the classical phase of an underlying, high-frequency oscillation (a strict analog to the usually-ignored rest mass oscillation at the Compton frequency). 

Despite the breadth of the above conclusions, this coupled-oscillator analog has a major drawback as an analog to an electron spin state.  It is limited by is the lack of coupling parameters that correspond to magnetic fields in the $x$ or $z$ directions, associated with the appropriate rotations around the Poincar\'e sphere.  The classical model in the next section solves this problem, although it comes at the expense of the Foucault pendulum's easily-visualized oscillations. 

\section{The Full Analog: Four Coupled Oscillators}

In order to expand the above example to contain an analog of an arbitrarily-directed magnetic field, two more coupling parameters must enter the classical Lagrangian.  But with only two oscillators, there are no more terms to couple.  With this in mind, one might be tempted to extend the above example to three coupled oscillators, but in that case the odd number of eigenmodes makes the dynamics unlike that of a spin-1/2 system.

It turns out that four coupled oscillators can solve this problem, so long as the eigenmodes come in degenerate pairs.  By extending $\bm{x}$ to a real 4-component vector (as opposed to the 2-component vector in the previous section), one can retain the same general form of the earlier Lagrangian: 
\begin{equation}
\label{L2}
L_2(\bm{x},\dot{\bm{x}},t)=\frac{1}{2}(\bm{p\cdot p}-\omega_0^2 \,\bm{x\cdot x}).
\end{equation}
Here we are still using the definition $\bm{p}\equiv\dot{\bm{x}}+\bm{B} \bm{x}$, but now with a 4x4 matrix encoding three independent coupling coefficients,
\begin{equation}
\label{eq:Bdef}
\bm{B} = \left[ \begin{array}{cccc}
0 & -\beta_z & \beta_y & -\beta_x \\
\beta_z & 0 & \beta_x & \beta_y \\
-\beta_y & -\beta_x & 0 & \beta_z \\
\beta_x  & -\beta_y & -\beta_z & 0 \end{array}  \right].
\end{equation}
Again, note that squaring the matrix $\bm{B}$ yields $\bm{B}^2=-\beta^2 \bm{I}$, where now $\beta \equiv \sqrt{\beta_x^2 + \beta_y^2 + \beta_z^2}$.  

\subsection{Constant Magnetic Fields}

The four corresponding Euler-Lagrange equations of motion (for constant $\beta$'s) can be written as
\begin{equation}
\label{eq:modes}
\left[ 2\bm{B}\frac{\partial}{\partial t}+\bm{I}\left(\omega_0^2-\beta^2+\frac{\partial^2}{\partial t^2} \right)  \right]\bm{x}=0.
\end{equation}
Solving (\ref{eq:modes}) for the eigenmodes via the replacement $\partial/\partial t \to -i\omega$ yields only two solutions, as the eigenvalues are doubly degenerate.  They are of the same form as in the previous section: $\omega_\pm = \omega_0 \pm \beta$.  

Because of the degeneracy, the full classical solutions can be expressed in a variety of ways.  It is convenient to consider a vector with the cartesian components $\bm{\beta}=(\beta_x,\beta_y,\beta_z)$, and then to transform it into spherical coordinates $(\beta,\theta,\phi)$.  Using the two-spinors $\ket{y_+}$ and $\ket{y_-}$ defined in (\ref{eq:defs}), the general solutions to (\ref{eq:modes}) can then be written as the real part of
\begin{align}
\label{eq:full}
\fspi{x_1(t)}{x_2(t)}{x_3(t)}{x_4(t)}= & \, a\spi{cos(\theta/2)\ket{y_-}}{sin(\theta/2)e^{i\phi}\ket{y_-}} e^{-i \beta t} + b\spi{sin(\theta/2)\ket{y_-}}{-cos(\theta/2)e^{i\phi}\ket{y_-}} e^{i \beta t} \notag \\
& + c\spi{-sin(\theta/2)e^{i\phi}\ket{y_+}}{cos(\theta/2)\ket{y_+}} e^{-i \beta t} + d\spi{cos(\theta/2)e^{i\phi}\ket{y_+}}{sin(\theta/2)\ket{y_+}} e^{i \beta t}.
\end{align}
Here the global $exp(-i\omega_0 t)$ dependence has been suppressed; one multiplies by this factor and takes the real part to get the actual coordinate values.  Having doubled the number of classical oscillators, the solution here is parameterized by {\it four} complex numbers ($a,b,c,d$).

This solution bears a striking similarity to the known dynamics of an electron spin state in an arbitrary uniform magnetic field $\vec{B}$ with components $(B,\theta,\phi)$.  In the basis defined above in (\ref{eq:defs}), those solutions to the SPE are known to be
\begin{equation}
\label{eq:qmev}
\spi{\chi_+ (t)}{\chi_- (t)} = f \spi{cos(\theta/2)}{sin(\theta/2)e^{i\phi}} e^{-ie Bt/2m} + g \spi{sin(\theta/2)}{-cos(\theta/2)e^{i\phi}} e^{ie Bt/2m},
\end{equation}
where the left side of this equation is the spinor $\ket{\chi (t)}$.  Here $f$ and $g$ are two complex constants subject to the normalization condition $|f|^2+|g|^2=1$.  

It is not difficult to see how all possible SPE solutions (\ref{eq:qmev}) have corresponding classical solutions (\ref{eq:full}).  Equating $\bm{\beta}=\vec{B}e/(2m)$, adding the quantum-irrelevant global phase dependence $exp(-i\omega_0 t)$ to $\ket{\chi (t)}$, and setting $c=d=0$ in (\ref{eq:full}) makes the two expressions appear almost identical if $a=\sqrt{2}f$ and $b=\sqrt{2}g$.  (The $\sqrt{2}$'s appear in the definition of $\ket{y_-}$).  The final step is to map the fully-real $x's$ to the complex $\chi's$ according to
\begin{equation}
\label{eq:map1}
\chi_+=x_1+ix_2 \, \, , \, \, \chi_- = x_3 + ix_4.
\end{equation}

This mapping turns out to be just one of many possible ways to convert a solution of the form (\ref{eq:qmev}) into the form (\ref{eq:full}).  For example, setting $a=b=0$, $c=\sqrt{2}f$ and $d=\sqrt{2}g$ corresponds to the alternate map
\begin{equation}
\label{eq:map2}
\chi_+=x_3-ix_4 \, \, , \, \, \chi_- = -x_1 + ix_2.
\end{equation}
More generally, one can linearly combine the above two maps by introducing two complex parameters $A$ and $B$.  Under the assignment $a=\sqrt{2}Af$, $b=\sqrt{2}Ag$, $c=\sqrt{2}Bf$ and $d=\sqrt{2}Bg$ (which can always be done if $ad\!=\!bc$) then the connection between the above equations (\ref{eq:full}) and (\ref{eq:qmev}) corresponds to
\begin{eqnarray}
\label{eq:map3}
\chi_+&=& \frac{(x_1+ix_2)A^*+(x_3-ix_4)B^*}{|A|^2+|B|^2}, \nonumber \\
\chi_- &=& \frac{(-x_1 + ix_2)B^*+(x_3 + ix_4)A^*}{|A|^2+|B|^2}.
\end{eqnarray}

This shows that for any solution (\ref{eq:full}) that obeys the $ad\!=\!bc$ condition, it will always encode a particular quantum solution to (\ref{eq:qmev}) via the map (\ref{eq:map3}), albeit with extra parameters $A$, $B$, and a specified global phase.  Remarkably, this $ad\!=\!bc$ condition happens to be equivalent to the simple classical constraint $L_2(t)=0$. Imposing such a constraint on (\ref{eq:modes}) therefore yields a classical system where all solutions can be mapped to the dynamics of a spin-1/2 quantum state in an arbitrary, constant, magnetic field -- along with a number of ``hidden variables'' not encoded in the quantum state.

\subsection{Time-varying Magnetic Fields}

As in Section III, a generalization to time-varying magnetic fields is best accomplished at the level of differential equations, not solutions.  Allowing $\bm{\beta}(t)$ to vary with time again adds a new term to the Euler-Lagrange equations, such that they now read:
\begin{equation}
\label{eq:ELx}
\left[ 2\bm{B}\frac{\partial}{\partial t}+\frac{\partial \bm{B}}{\partial t}+\bm{B}^2+\bm{I}\left(\omega_0^2+\frac{\partial^2}{\partial t^2} \right)  \right]\bm{x}=0.
\end{equation}
Here $\bm{B}$ is given by (\ref{eq:Bdef}) with time-dependent $\beta_x$, $\beta_y$, and $\beta_z$.  This must be again compared with the SPE with an explicit  rest mass oscillation $mc^2=\hbar\omega_0$:
\begin{equation}
\label{eq:SE}
i\hbar \frac{\partial}{\partial t} \spi{\chi_+}{\chi_-} = \hbar \left( \omega_0 \bm{I}+\bm{\beta}\cdot\bm{\sigma} \right) \spi{\chi_+}{\chi_-},
\end{equation}
where again we have used $\bm{\beta}(t)=\vec{B}(t)e/(2m)$ to relate the coupling parameters in $L_2$ with the magnetic field $\vec{B}(t)$.  (Here $\bm{\sigma}$ is the standard vector of Pauli matrices).

While it is possible to use the map (\ref{eq:map3}) to derive (\ref{eq:ELx}) from (\ref{eq:SE}) (and its time-derivative) via brute force, it is more elegant to use the quaternion algebra, as it is closely linked to both of the above equations.  Defining the two quaternions $\mathfrak{q}=x_1+ix_2+jx_3+kx_4$, and $\mathfrak{b}=0+i\beta_z-j\beta_y+k\beta_x$, allows one to rewrite (\ref{eq:ELx}) as the quaternionic equation
\begin{equation}
\label{eq:ELq}
\ddot{\mathfrak{q}}+2\dot{\mathfrak{q}}\mathfrak{b}+\mathfrak{q}(\mathfrak{b}^2+\dot{\mathfrak{b}}+\omega_0^2)=0.
\end{equation}
Note that while $\bm{B}$ operates on $\bm{x}$ from the left, the $\mathfrak{b}$ acts as a right-multiplication on $\mathfrak{q}$ because (\ref{eq:Bdef}) is of the form of a right-isoclinic rotation in SO(4). 

While it is well-known that the components of $i\bm{\sigma}$ act like purely imaginary quaternions, the precise mapping of $i\bm{\beta}\cdot\bm{\sigma}$ to $\mathfrak{b}$ depends on how one maps $\ket{\chi}$ to a quaternion $\mathfrak{s}$.  Using the above map (\ref{eq:map1}), combined with the above definition of $\mathfrak{q}$, it happens that $i(\bm{\beta}\cdot\bm{\sigma})\ket{\chi}=\mathfrak{s}\mathfrak{b}$, where $\mathfrak{s}$ is the quaternionic version of $\ket{\chi}$ (as defined by the combination of (\ref{eq:map1}) and $\mathfrak{q}=\mathfrak{s}$).  This allows one to write the SPE (\ref{eq:SE}) as
\begin{equation}
\label{eq:SEq}
\dot{\mathfrak{s}}+\mathfrak{s}\mathfrak{b}=-i\omega_0\mathfrak{s}.
\end{equation}
This equation uses a quaternionic $i$, not a complex $i$, acting as a left-multiplication (again because of the particular mapping from $\ket{\chi}$ to $\mathfrak{s}$ defined by (\ref{eq:map1})).  While the SPE would look more complicated under the more general map (\ref{eq:map3}) as applied to $\mathfrak{q}=\mathfrak{s}$, this is equivalent to applying the simpler map (\ref{eq:map1}) along with
\begin{equation}
\label{eq:qtos}
\mathfrak{q}=[Re(A)+iIm(A)+jRe(B)-kIm(B)]\mathfrak{s}\equiv \mathfrak{u}\mathfrak{s},
\end{equation}
so long as $\mathfrak{u}$ is a constant unit quaternion (linking the normalization of $\mathfrak{q}$ and $\mathfrak{s}$).  

Keeping the SPE in the form (\ref{eq:SEq}), we want to show that for any solution $\mathfrak{s}$ to (\ref{eq:SEq}), there is a family of solutions $\mathfrak{q=us}$ to the classical oscillators (\ref{eq:ELq}).  The time-derivative of (\ref{eq:SEq}) can be expanded as
\begin{equation}
\label{eq:SEq2a}
\ddot{\mathfrak{s}}+2\dot{\mathfrak{s}}\mathfrak{b}+\mathfrak{s}\dot{\mathfrak{b}} = \dot{\mathfrak{s}}\mathfrak{b}-i\omega_0\dot{\mathfrak{s}}.
\end{equation}
Using (\ref{eq:SEq}) to eliminate the $\dot{\mathfrak{s}}$'s on the right side of (\ref{eq:SEq2a}) then yields
\begin{equation}
\label{eq:SEq2}
\ddot{\mathfrak{s}}+2\dot{\mathfrak{s}}\mathfrak{b}+\mathfrak{s}(\mathfrak{b}^2+\dot{\mathfrak{b}}+\omega_0^2)=0.
\end{equation}
If $\mathfrak{s}$ solves (\ref{eq:SEq}), it must solve (\ref{eq:SEq2}), but this is exactly the same equation as (\ref{eq:ELq}).  And because $\mathfrak{u}$ is multiplied from the left, $\mathfrak{q=us}$ must then also solve (\ref{eq:ELq}).  This concludes the proof that all solutions to the SPE (\ref{eq:SE}) -- even for a time-varying magnetic field -- have an exact classical analog in the solutions to (\ref{eq:ELx}).  

The question remains as to which subset of solutions to (\ref{eq:ELx}) has this quantum analog.  If the above connection exists between $\mathfrak{q}$ and $\mathfrak{s}$, then by definition $\mathfrak{s=u^*q}$, where $\mathfrak{u}$ is a unit quaternion.  This substitution transforms the left side of (\ref{eq:SEq}) into $\mathfrak{u^*p}$, where $\mathfrak{p}=\dot{\mathfrak{q}}+\mathfrak{qb}$ is the quaternionic version of the canonical momentum.  Therefore, from (\ref{eq:SEq}), $\mathfrak{p}=-\mathfrak{u}i\omega_0\mathfrak{u^*q}$.  As $\mathfrak{u}$ is a unit quaternion, this yields a zero Lagrangian density $L_2=(|\mathfrak{p}|^2-\omega_0^2|\mathfrak{q}|^2)/2=0$, consistent with the constant-field case.

\section{Discussion}

The Foucault pendulum is often discussed in the context of classical analogs to quantum spin states\cite{Klyshko}, but the discussion is typically restricted to geometric phase.  Section III demonstrated that the analogy runs much deeper, as the Coriolis coupling between the two oscillatory modes is exactly analogous to a one-dimensional magnetic field acting on an electron spin state.  The analog also extends to the dynamics, and provides a classical description of Zeeman energy splitting, geometric phase shifts, and the appearance of a doubled gyromagnetic ratio.  Apart from a global phase, there were no additional classical parameters needed to complete the Section III analog.  

In Section IV, we demonstrated that it is possible to take four classical oscillators and physically couple them together in a particular manner (where the three coupling coefficients correspond to the three components of a magnetic field), yielding the equations of motion given in (\ref{eq:ELx}).  Imposing a global physical constraint ($L_2=0$) on this equation forces the solutions to have an exact map (\ref{eq:map3}) to solutions of the Schr\"odinger-Pauli equation for a two-level quantum system with a rest-mass oscillation.  This is a many-to-one map, in that there are additional parameters in the solution to (\ref{eq:ELx}) that can be altered without affecting the corresponding quantum solution, including an overall phase.  From a quantum perspective, these additional parameters would be ``hidden variables''.

Perhaps one reason this analog has not been noticed before is that many prior efforts to find classical analogs for the spin-1/2 state have started with a physical angular momentum vector, in real space.  Rotating such a physical vector by $2\pi$, it is impossible to explain a $\pi$ geometric phase shift without reference to additional elements outside the state itself, such as in Feynman's coffee cup demonstration \cite{Feynman}.  In the four-oscillator analog, however, the expectation value of the spin angular momentum merely corresponds to an unusual combination of physical oscillator parameters:
\begin{equation}
\label{eq:expS}
\left< \bm{S\cdot\hat{e}} \right> = -\frac{\hbar}{2\omega_0} \bm{p\cdot B(\hat{e}) x}.
\end{equation}
Here $\bm{\hat{e}}$ is an arbitrary unit vector, and the above definition of $\bm{B(\beta)}$ in (\ref{eq:Bdef}) is used to define $\bm{B(\hat{e})}$.  Note, for example, that a sign change of both $\bm{x}$ and $\bm{p}$ leaves $\left<\bm{S}\right>$ unchanged.  This is indicative of the fact that the overall phase of the oscillators are shifted by $\pi$ under a $2\pi$ rotation of $\left<\bm{S}\right>$, exactly as in the CP-analog and the Foucault pendulum.  

This result explicitly demonstrates that if there is any inherently non-classical aspect to a quantum spin-1/2 state, such an aspect need not reside in the dynamics.  On the other hand, if the system is measured, this classical analog cannot explain why superpositions of eigenmodes are never observed, or indeed what the probability distribution of measurements should be.  That analysis resides in the domain of quantum measurement theory, and these results do not indicate whether or not that domain can be considered  to have a classical analog.

With this in mind, these results should still be of interest to approaches where the usual quantum state is not treated as a complete description of reality.  The hidden variables that naturally emerge from the above analysis are the complex parameters $A$ and $B$ (or equivalently, the unit quaternion $\mathfrak{u}$).  These parameters effectively resulted from the doubling of the parameter space (from two to four oscillators), but do not seem to have any quantitative links to prior hidden-variable approaches.  Still, they are loosely aligned with the doubling of the ontological state space in Spekkens's toy model \cite{Spekkens}, as well as with the doubling of the parameter space introduced when moving from the first-order Schr\"odinger equation to the second-order Klein Gordon equation \cite{KGE}.  Another point of interest is that this analog stems from a relatively simple Lagrangian, $L_2$, and there is good reason to believe than any realistic model of quantum phenomena should have the same symmetries as a Lagrangian density \cite{WMP}. 

One final question raised by these results is whether or not it is possible to construct a working mechanical or electrical version of the classical oscillators described in Section IV.  If this were possible, it would make a valuable demonstration concerning the dynamics of an unmeasured electron spin state.  Even if it were not possible, some discussion of these results in a quantum mechanics course might enable students to utilize some of their classical intuition in a quantum context.

\begin{acknowledgments}
The authors are indebted to Patrick Hamill for recognizing (\ref{eq:L1}) as the Foucault pendulum Lagrangian; further thanks are due to Ian Durham, David Miller, and William Wharton.  An early version of this work was completed when KW was a Visiting Fellow at the Centre for Time in the Sydney Centre for Foundations of Science, University of Sydney.

\end{acknowledgments}

\end{document}